\documentclass[11pt]{article}
\usepackage{moriond,epsfig}

\newlength{\picwi}
\picwi\textwidth

\begin{document}
\vspace*{4cm}
\title{READINESS OF THE ATLAS EXPERIMENT FOR FIRST DATA}

\author{T. PAULY\\ on behalf of the ATLAS Collaboration}

\address{CERN, PH Department, Route de Meyrin, \\
1211 Geneva 3, Switzerland}

\maketitle\abstracts{The ATLAS detector is one of the experiments at the LHC that will detect high-energy proton collisions at 14\,TeV. The commissioning of the detector has started already in 2005 in parallel to the detector installation and is still in progress. The data taken so far corresponds to noise runs, cosmic muon events and beam background events from single beam in September 2008. We present the current status of the detector and performance results obtained during commissioning.}

\section{Introduction}
The ATLAS experiment is a general purpose detector built to study high-energy proton-proton collisions at the LHC and is described in detail elsewhere\cite{bib:atlas}.
Commissioning of the detector has started in 2005 in parallel to its installation.
Large samples of cosmic muons have been recorded, which are used to understand and improve the performance of the detector, in particular for detector alignment and for first calibrations.
The data are very useful to test channel mappings and the timing,  to determine dead and noisy channels, and to verify the stability of the hardware during operation.
In addition, they help in gaining experience in the detector operation, the data acquisition and the analysis chain.
Data was also recorded in September 2008 with first single beams circulating in the LHC.

\section{Inner Detector}
The Inner Detector consists of the Pixel detector, the Semiconductor Tracker (SCT) and the Transition Radiation Tracker (TRT), which are operated inside a 2\,T magnetic solenoid field parallel to the beam axis.
It has a coverage in pseudo-rapidity $\eta$ of $|\eta|<2.5$ (TRT $|\eta|<2$) and a momentum resolution of $\sigma/p_T = 0.05 \%\cdot p_T/\mathrm{GeV} \oplus 1 \%$.

The Pixel Detector consists of 3 layers in the barrel and end-cap regions, with 80 million pixels of size $50\,\mu$m$\times 400\,\mu$m.
More than 95\,\% of the modules are operational, the noise occupancy was measured to be about $5\times 10^{-9}$ and the hit efficiency to be better than 98\,\%.

\begin{figure}[pht]
\begin{center}
\includegraphics[width=0.5\picwi]{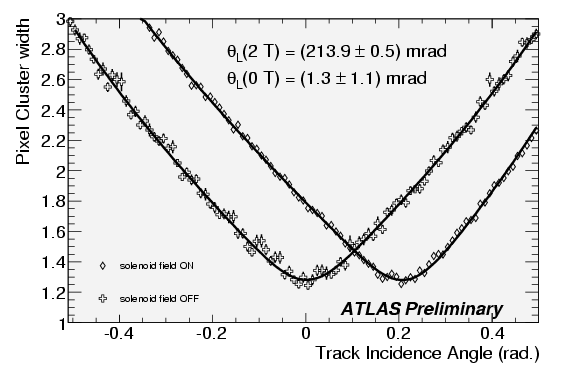}
\caption{\label{fig:pix-lorentz}Cluster width versus track incidence angle for pixel clusters on tracks.}
\end{center}
\end{figure}
Figure\,\ref{fig:pix-lorentz} shows width versus angle of incidence in azimuth for pixel clusters on tracks.
Shown are both the data for the solenoid off (left curve) and the solenoid on (right curve).
The minimum of this distribution determines the Lorentz angle and the fit result is shown.
As expected, the Lorentz angle is consistent with 0 for the data without magnetic field.
Using the data with magnetic field the preliminary measurement of the Lorentz angle is ($213.9 \pm 0.5$)\,mrad; the expectation is about 225\,mrad.

The SCT consists of 4 double layers of 80\,$\mu$m thin silicon strips in the barrel region and 9 in each end-cap.
It has a total of 6 million channels spread over 4088 modules.
More than 99\,\% of the modules are operational in the barrel and more than 97\,\% in the end-caps.
The noise occupancy was measured to be $4.4\times 10^5$ for the barrel and $5\times 10^5$ for the end-caps.
The single hit efficiency is larger than 99\,\%.

\begin{figure}[pht]
\begin{center}
\parbox{0.47\picwi}{
\includegraphics[width=0.45\picwi]{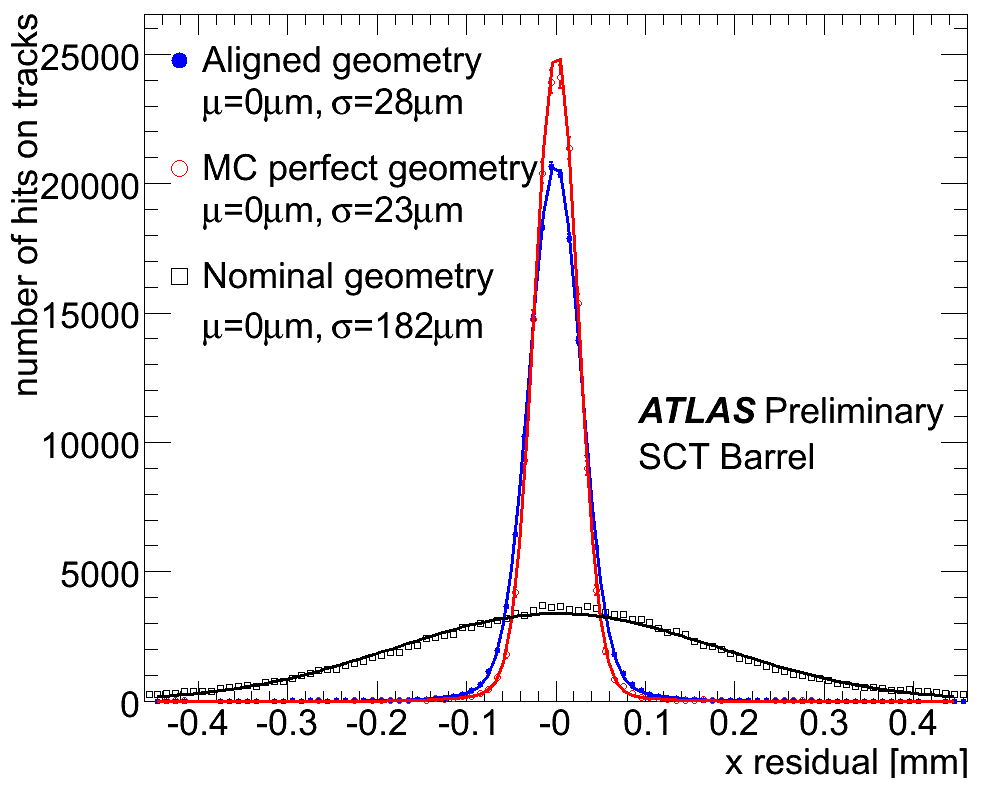}
\caption{\label{fig:sct-residual}Track residuals in the precision coordinate of the SCT barrel.}
}
\parbox{0.47\picwi}{
\includegraphics[width=0.45\picwi]{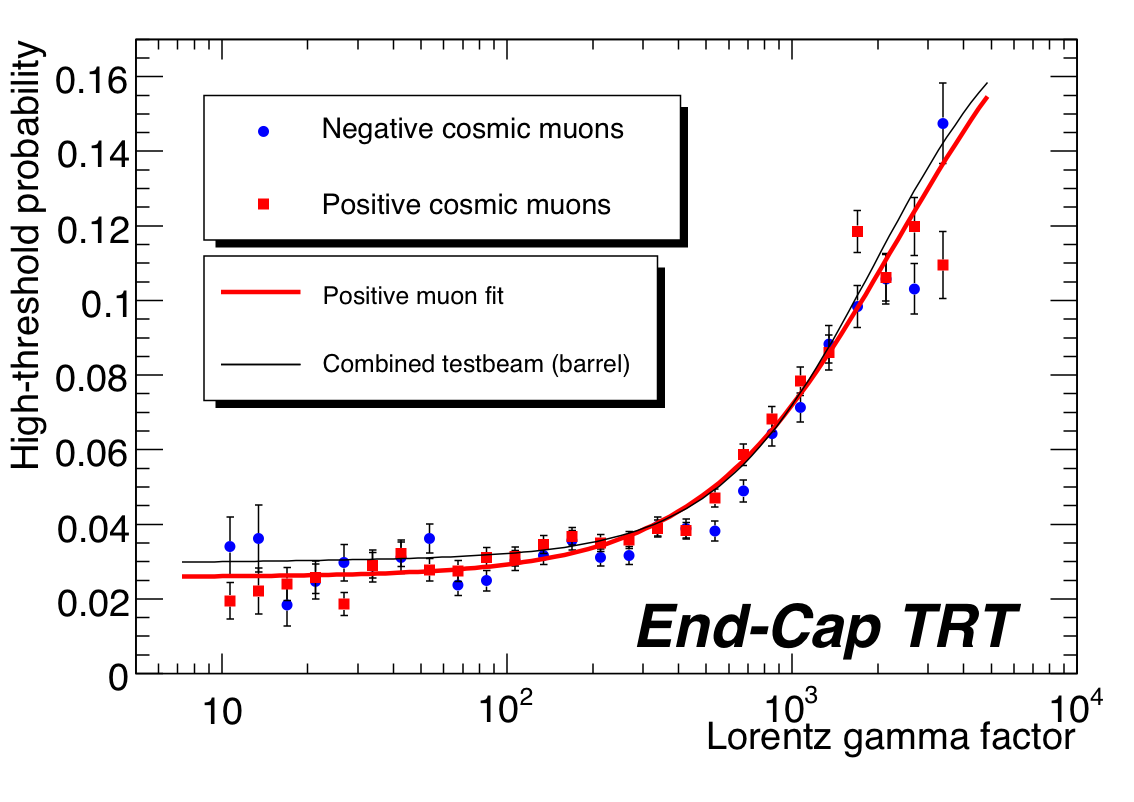}
\caption{\label{fig:trt}Probability for high-threshold hits -- an indicator for transition radiation -- versus the Lorentz gamma factor for cosmic muons.}
}
\end{center}
\end{figure}
The residual distribution in the precision coordinate, integrated over all hits-on-tracks, can be seen in Figure\,\ref{fig:sct-residual} for the SCT barrel, for the nominal and the preliminary aligned geometry.
The residual is defined as the measured hit position minus the expected hit position from the track extrapolation.
Shown is the projection onto the local $x$-coordinate -- the precision coordinate.
A single Gaussian fit gives a resolution of 28\,$\mu$m, which is already close to the expectation for the perfect geometry (23\,$\mu$m).
Similar results are available for the Pixel barrel $x$-direction (23\,$\mu$m), Pixel barrel $y$-direction (134\,$\mu$m), and TRT barrel (174\,$\mu$m).

The TRT is a combined straw tube tracker and transition radiation detector, which allows electron-pion identification in the energy region between 500\,MeV and 150\,GeV.
The straw tubes are 4\,mm in diameter and contain a 35\,$\mu$m thin anode wire.
The barrel region consists of 73 layers of axial straws and the end-cap regions of 160 layers of radial straws per end-cap.
About 98\,\% of the channels are operational.

%In the TRT, TR photons are detected by absorption of the photons in the chamber gas (Xenon mixture, short absorption length for photons) leading to high electronic pulses crossing a high threshold (pulses from particles which do not produce TR usually do only cross a low threshold).
%On the plot one can see the turn-on of the production of TR photons as a function of gamma as measured for the tracks of cosmic particles (Muons) during cosmic data taking of the ATLAS detector in October 2008. On the y-axis the probability of a high-threshold hit (indicator for TR) is given.
Figure\,\ref{fig:trt} shows the probability of high-threshold hits -- an indicator for the production of transition radiation photons -- as a function of the relativistic gamma factor measured for tracks from cosmic muons.
The data points are shown for both muon charges (positive: red squares, negative: blue dots) and are compared to the results obtained in the ATLAS Combined Test Beam in 2004 (black line).
The red line corresponds to a fit to the results obtained with the cosmic data.
The turn-on of the transition radiation is clearly visible and the detector responds identically to cosmic tracks and data recorded at the test beam.
%TRTBarrelResidual.png

\section{Calorimeters}
ATLAS includes two types of sampling calorimeters: the Liquid Argon Calorimeter and the Tile Calorimeter.
The Liquid Argon Calorimeter comprises the electromagnetic calorimetry in the barrel and end-cap regions, as well as the hadronic calorimetry in the end-cap and forward regions.
The electromagnetic part consists of layers of lead and liquid argon in accordion geometry, with 3 longitudinal samples in the region of $|\eta| < 2.5$.
A pre-shower detector assists in $|\eta|<1.8$.
The hadronic calorimeter in the end-cap regions uses copper as absorber material and has 4 longitudinal samples, whereas in the forward region, tungstate and 3 longitudinal samples are used.
The electromagnetic energy resolution is $\sigma/E = 10\,\%/\sqrt{E/\mathrm{GeV}} \oplus 0.7\,\%$ and the hadronic energy resolution $\sigma/E = 100\,\%/\sqrt{E/\mathrm{GeV}} \oplus 10\,\%$ for $3.1<|\eta|<4.9$.
About 0.02\,\% of dead channels were found, in addition to 0.9\,\% of dead but recoverable channels.
There are about 0.003\,\% of noisy channels.
% and less than 0.2\,\% of channels with bad calibration.
The electronic calibration procedure is operational and the calibration constants are used online.

\begin{figure}[pht]
\begin{center}
\parbox{0.45\picwi}{
\includegraphics[width=0.43\picwi]{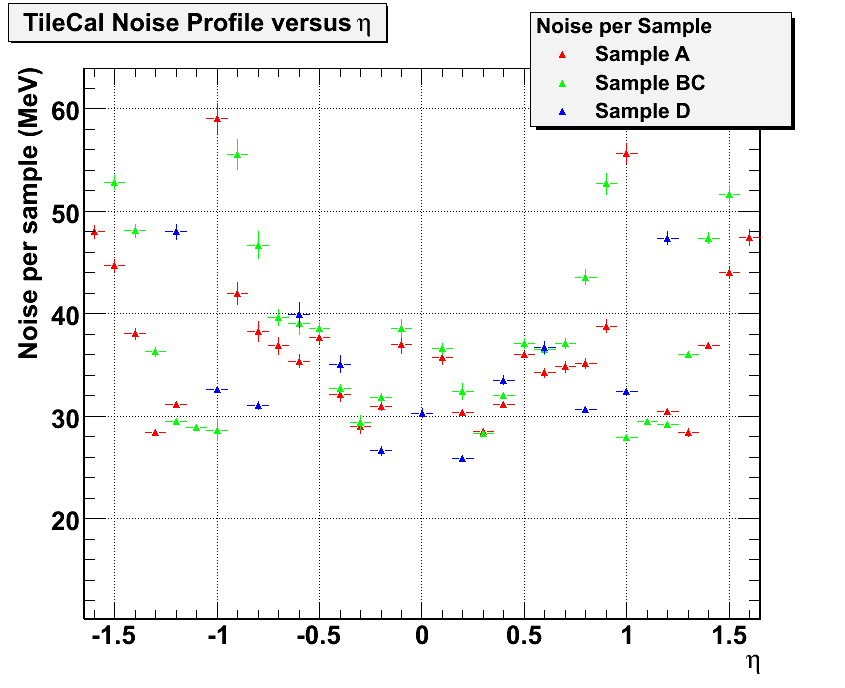}
\caption{\label{fig:tile1} Noise distribution in the Tile Calorimeter during first LHC beam.}
}
\parbox{0.54\picwi}{
\includegraphics[width=0.50\picwi]{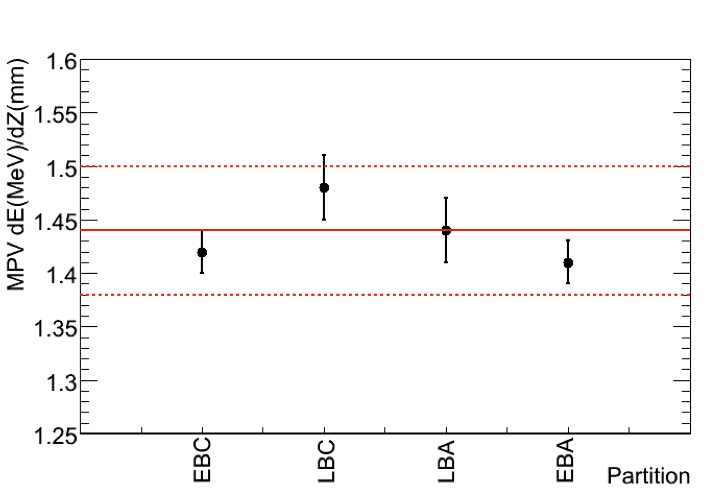}
\caption{\label{fig:tile2} Inter-calibration of the four Tile Calorimeter cylinders for events from first LHC beam.}
}
\end{center}
\end{figure}
The Tile Calorimeter performs the hadronic calorimetry in the barrel region.
It consists of scintillator tiles and iron absorber with 3 longitudinal samples.
The hadronic energy resolution is $\sigma/E = 50\,\%/\sqrt{E/\mathrm{GeV}} \oplus 3\,\%$ for $|\eta|<3.2$.
The fraction of currently dead, but to be repaired channels is less than 1.4\,\%.
All calibration systems, based on a Cs source, laser, and charge injection, are operational.

The Level-1 Calorimeter Trigger processor uses trigger towers (usually $0.1\times 0.1$ in pseudo-rapidity and azimuthal angle) from the two calorimeters to reconstruct electron/photon and tau/hadron clusters, jets, missing energy and energy sums for the Level-1 trigger system.
Out of 7200 analogue channels, less than 0.4\,\% are dead, in addition to 0.3\,\% of recoverable channels.
Channel-to-channel noise suppression currently allows a cut on transverse energy  as low as 1\,GeV (aim: 0.5\,GeV).

\begin{figure}[pht]
\begin{center}
\parbox{0.49\picwi}{
\includegraphics[width=0.45\picwi]{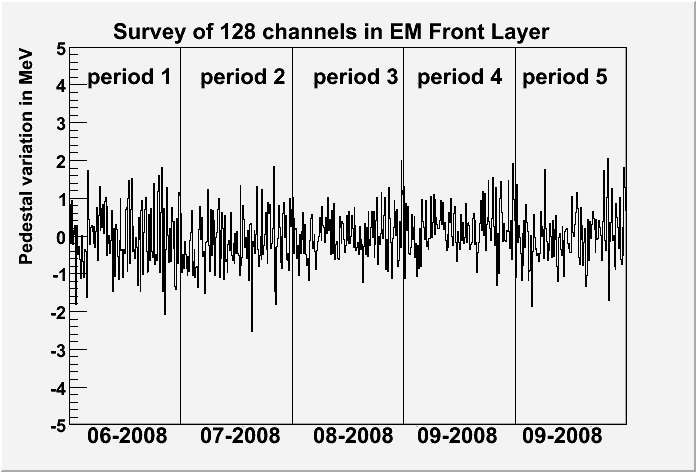}
\caption{\label{fig:lar1}Liquid Argon Calorimeter pedestal variation for 128 channels of the electromagnetic front layer over a period of 5 months.}
}
\parbox{0.43\picwi}{
\includegraphics[width=0.39\picwi]{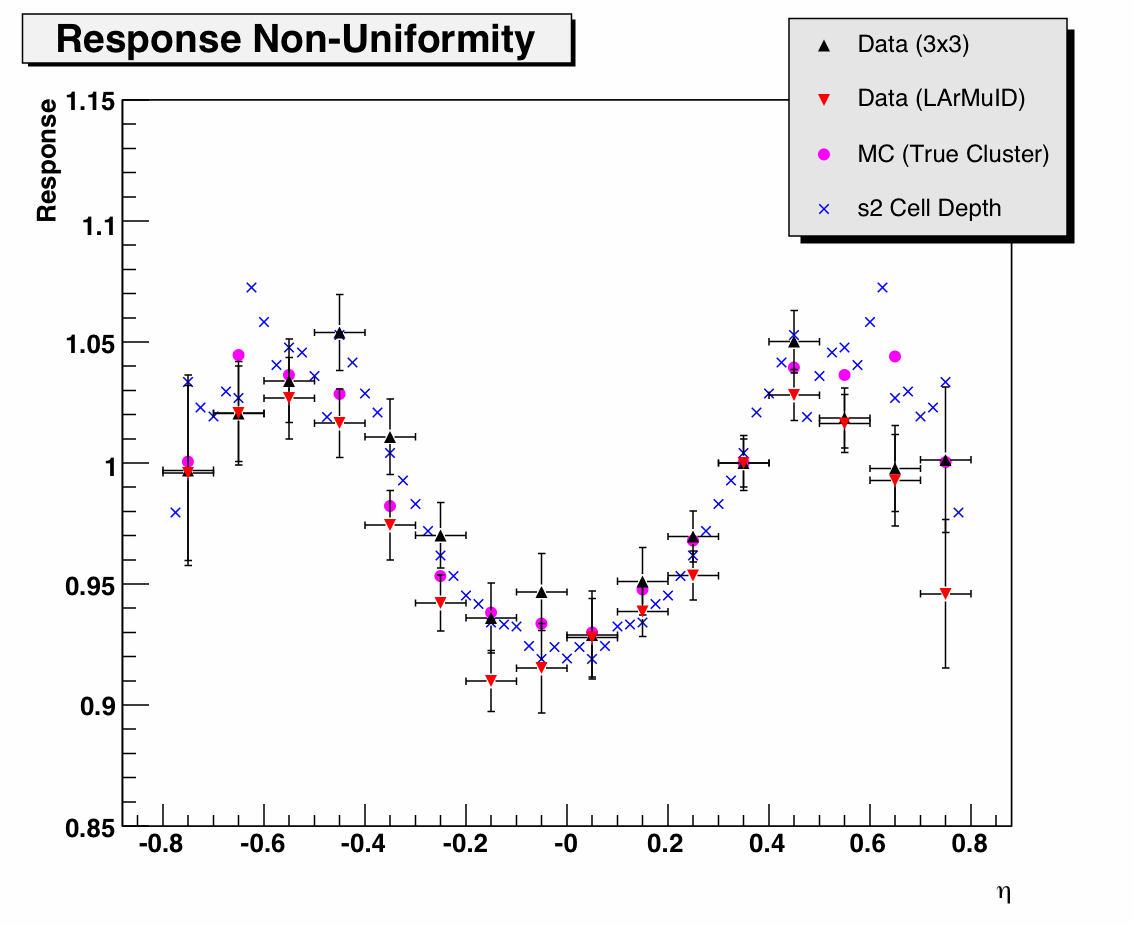}
\caption{\label{fig:lar2}Response of the Liquid Argon Ca\-lo\-ri\-me\-ter to minimum ionising particles as a function of pseudo-rapidity.}
}\end{center}
\end{figure}
Figure\,\ref{fig:tile1} shows the noise distribution for the Tile calorimeter, measured during the period of first LHC beam in September 2008 with events triggered by a random trigger.
The root mean squared of the cell energy distribution is displayed against the pseudo-rapidity of the calorimeter cells, for cells belonging to different radial samples.
The results are compatible with the noise level measured during cosmic-ray data taking.
%Tile Calorimeter average electronic noise measured in Sep. 2008, during LHC first beam period. Events have been triggered by RNDM trigger. The pseudorapidity (eta) of the cells is shown on the Xaxis, while the Yaxis represent the RMS of the cells energy distribution. The error bar represent the spread of the noise of cells with the same pseudorapidity during Sep. 2008. Cells belonging to different radial samples of the Tile Calorimeter have been represented with different colors. Results are compatible with the noise level measured during cosmic data taking. 
Figure\,\ref{fig:tile2} shows the average of the most probable energy loss d$E$/d$x$ recorded with horizontal muons from single beam data (September 10, 2008) for the four calorimeter cylinders.
The inter-calibration of the cylinders is within 4\,\%, as expected from the calibration with radioactive gamma sources.
%Most Probable Value of dE/dx signals recorded by TileCal with horizontal muons from single beam data on Sept. 10, 2008. The average over all cells within a given partition response to horizontal muons is shown for each partition. About 500 muons were selected by requiring consistent to expected signal along 12m of Tile calorimeter length. This data provided the opportunity to verify the intercalibration of Tile calorimeter cylinders, already calibrated with radioactive gamma sources, down to the 4\,\% precision level. The red lines represent the average MPV value of the 4 barrels and its 4% uncertainty. 

\begin{figure}[pht]
\begin{center}
\parbox{0.51\picwi}{
\includegraphics[width=0.49\picwi]{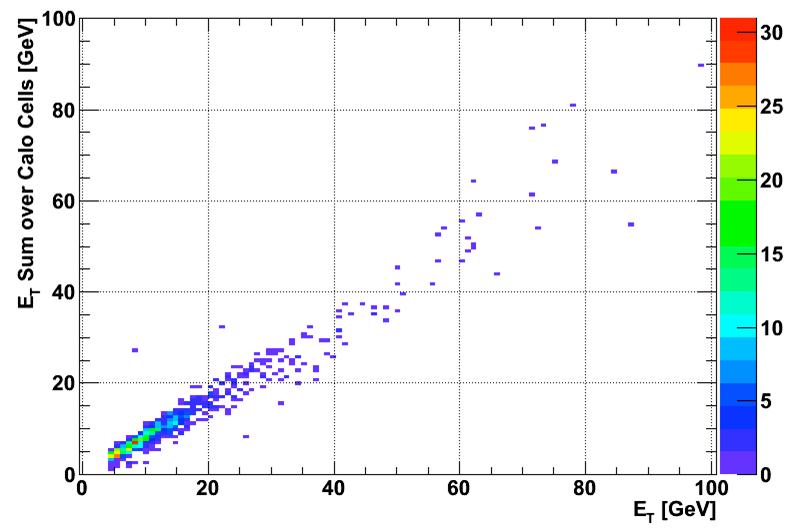}
\caption{\label{fig:l1calo1} Correlation between the energy reconstructed by the Level-1 Calorimeter Trigger processor (horizontal axis) and the full precision read-out of the Liquid Argon electromagnetic layers (vertical axis).}
}
\parbox{0.48\picwi}{
\includegraphics[width=0.42\picwi]{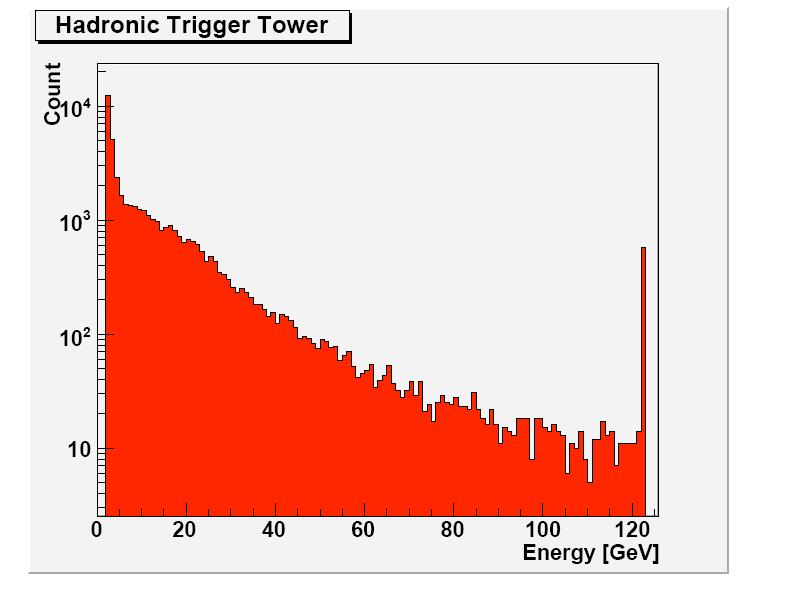}
\caption{\label{fig:l1calo2} Energy spectrum of trigger towers reconstructed by the Level-1 Calorimeter Trigger processor for a typical cosmic run.}
}
\end{center}
\end{figure}
\newpage
For the Liquid Argon Calorimeter, the pedestal variation in MeV for 128 channels of the electromagnetic front layer is displayed in Figure\,\ref{fig:lar1}.
The variations are stable within $\pm 1$\,MeV over a period of five months.
In Figure\,\ref{fig:lar2}, the response non-uniformity of the Liquid Argon Calorimeter can be seen: the most probable energy extracted from a fit to the energy distribution as a function of pseudo-rapidity.
Curves of two different clustering algorithms are shown, along with the results from a simulation and the cell depth of the second sampling.
The two distributions clearly track the cell depth as expected for a minimum ionising particle, and the uniformity of the response agrees with the simulation at the level of 2\,\%.

Figure\,\ref{fig:l1calo1} shows, for cosmic events, the correlation between the energy reconstructed in trigger towers by the Level-1 Calorimeter Trigger processor (horizontal axis) and the energy measured in the full readout for the Liquid Argon electromagnetic layers (vertical axis).
A good correlation is seen, even with preliminary calibrations.
%
%L1Calo calorimeter energy correlation plot. This shows the energy measured in trigger towers (usually 0.1x0.1 in eta/phi) by the Level-1 calorimeter trigger (x-axis) compared to the energy measured in the full readout for the calorimeters (y-axis) - in this case the Liquid Argon electromagnetic layers. A good correlation is seen, and this is done with nominal calibrations. When full energy calibration is applied, the spread will decrease. A sharp cut-off is seen in the trigger tower energy at the trigger threshold of 5 GeV used for much of the 2008 cosmic running.
In Figure\,\ref{fig:l1calo2}, we see for a typical cosmic run, the energy spectrum of trigger towers as reconstructed in the Level-1 Calorimeter trigger towers.
Though the majority of cosmic muons only deposit a small amount of energy in the calorimeters, there is a large tail of higher energies, mainly due to showering and large showers, rather than individual muons.
The peak at the top end of the spectrum is an artifact of saturation in the digital logic for the setting used at the time -- all energies above about 125\, GeV are recorded (and triggered) as saturated towers.
%l1calospectrum.JPG
%L1Calo cosmic spectrum. Here we see the energy spectrum of trigger tower energies as seen in the Level-1 Calorimeter trigger towers for a typical cosmic run. Though the majority of cosmic muons only deposit a small ammount of energy in the calorimeters, there is a large tail of higher energies, mainly due to showering and large showers, rather than individual muons. These are very useful for debugging the trigger. The peak at the top end of the spectrum is an artefact of saturation in the digital logic for the setting used at the time - all energies above about 125 GeV are recorded (and triggered) as saturated towers.

\begin{figure}[pht]
\begin{center}
\parbox{0.42\picwi}{
\includegraphics[width=0.4\picwi]{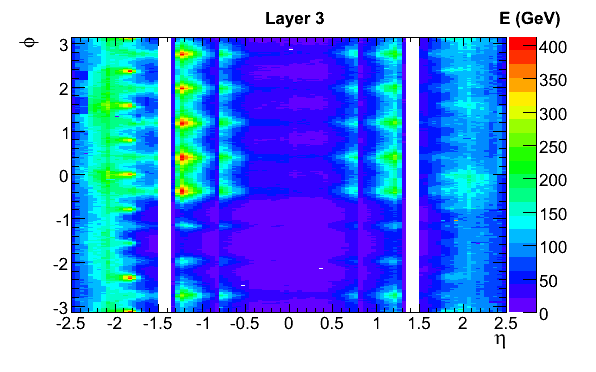}
\caption{\label{fig:calo-splash1}Energy deposits in the Liquid Argon Calorimeter for a 'beam-splash' event.}
}
\parbox{0.57\picwi}{
\includegraphics[width=0.55\picwi]{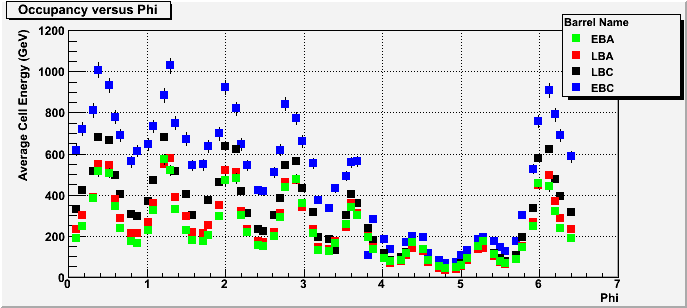}
\caption{\label{fig:calo-splash2}Energy deposits in the Tile Calorimeter for a 'beam-splash' event.}
}
\end{center}
\end{figure}
Figure\,\ref{fig:calo-splash1} and \ref{fig:calo-splash2} show the energy deposit in the calorimeters for events where a low intensity LHC bunch was intentionally hitting a closed collimator 140\,m upstream of ATLAS, resulting in a large particle shower, a so-called 'beam-splash'.
The energy of the third layer of the Liquid Argon electromagnetic calorimeter in pseudo-rapidity and azimuthal angle is shown in Figure\,\ref{fig:calo-splash1}, whereas Figure\,\ref{fig:calo-splash2} displays the average cell energy versus azimuthal angle for the Tile calorimeter.
A convolution of two effects is visible: the profile of the tunnel, whose floor at azimuth $-\pi/2$ or $3\pi/2$ absorbs a large fraction of the particles, and the 8-fold structure in azimuth introduced by the material of the end-cap toroid magnet in front of the calorimeters.
%This picture shows the 8-fold structure in phi of the beam splash events recorded the 10th of September. The structure is due to the End Cap toroid material in front of TileCal for particles coming from the C-side. The up -down asymmetry is also due to the material in fron of the detector 

\section{Muon Spectrometer}
The muon spectrometer consists of 4 different kinds of muon chambers covering the region of $|\eta|<2.5$ and an air-core toroid magnet system of rigidity 1.5-5.5\,T\,m for $|\eta|<1.4$ and 1-7.5\,T\,m for $|\eta|>1.6$.
The expected standalone momentum resolution is $\sigma/p_T < 10\,\%$ up to 1\,TeV.

For triggering at Level-1, fast chambers are used with a time resolution below 10\,ns and 2-dimensional readout with a space resolution of 5-10\,mm.
The barrel region consists of 544 resistive-plate chambers (RPC) with 359 thousand channels.
About 70\,\% of the chambers are currently operational, with the goal of having 95.5\,\% operational by summer 2009.
The fraction of dead strips is below 2\,\% and the fraction of hot strips and spots below 1\,\%.
In the end-cap region, 3588 thin-gap chambers (TGC) are used with 318 thousand channels.
About 99.8\,\% of the chambers are operational, dead channels are less than 0.01\,\% and noisy channels (with an occupancy higher than 5\,\%) are below 0.02\,\%.

For precision tracking, precision chambers are used with a spatial resolution of 35-40\,$\mu$m.
The barrel and end-cap regions are instrumented with 1088 stations of monitored drift tubes (MDT) with 339 thousand channels.
The stations are aligned with an optical alignment system that consists of 12232 sensors.
About 99.8\,\% of the chambers are operational, with 0.1\,\% dead channels and an additional 1\,\% of recoverable channels.
The fraction of noisy channels with at least 5\,\% occupancy is less than 0.2\,\%.
In the forward direction, 32 cathode-strip chambers (CSC) are used, with 31 thousand channels.
All of the chambers are operational and the fraction of dead channels is less than 0.1\,\%.

\begin{figure}[pht]
\begin{center}
\parbox{0.47\picwi}{
\includegraphics[width=0.45\picwi]{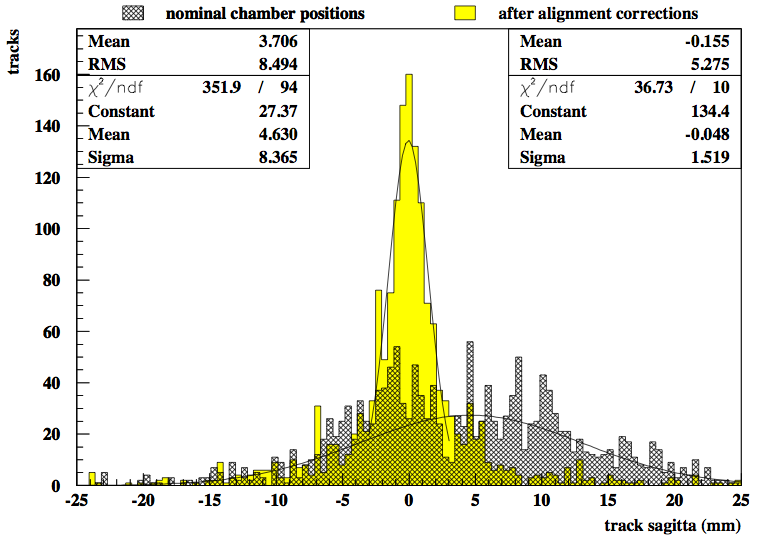}
\caption{\label{fig:mdt-sagitta}Sagitta distribution for cosmic muon tracks reconstructed by the MDT, before and after optical alignment.}
}
\parbox{0.47\picwi}{
\includegraphics[width=0.45\picwi]{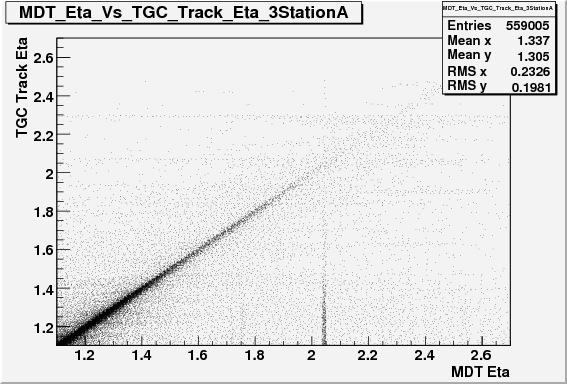}
\caption{\label{fig:mdt-tgc}Correlation of the pseudo-rapidity coordinate of tracks reconstructed by the TGC (vertical axis) and the MDT (horizontal axis).}
}
\end{center}
\end{figure}
Figure\,\ref{fig:mdt-sagitta} illustrates the performance of the optical alignment system for one station of monitored drift tubes.
The sagitta -- fake due to chamber misalignment -- of straight muon tracks is shown before and after optical alignment.
After alignment, a mean position of $(48\pm 54)\,\mu$m is observed, which is consistent with zero and the claimed accuracy of the optical alignment system of 45-50\,$\mu$m.
The width of 1.5\,mm is compatible with the expected amount of multiple scattering.
%The plot shows "track" sagittas calculated from segments in the stations EI-EM-EO before and after applying alignment corrections determined by the optical alignment system of the end-caps. We do not actually use any of the tracking algorithms, instead we define as a "track" a triplet of segments that have passed our selection cuts. The sagitta is then calculated in the usual way as the distance in the precision coordinate of the EM segment from the line joining the EI-EO segments.
%The "nominal chamber positions" histogram reflects the positioning accuracy that we have reached in the end-caps - typically 5mm rms in all coordinates for the chambers within a wheel, but up to 25mm displacements of entire wheels along ATLAS-Z. The EM station is displaced significantly along ATLAS-Z on the A-side, but not on the C-side, which is the main contribution to the difference between sides A and C.
%The "after alignment corrections" histogram should, for the optical alignment to be ok, be centered at zero and have a width that can be explained by effects other than random chamber misalignment. We observe a mean value of -48 +/- 54mu, thus compatible with zero, and a width of 1.5mm, compatible with the expected multiple-scattering width (as a benchmark, for 40GeV muons at theta=30 degrees the expected width from a single-muon simulation is 1.4mm).
%We conclude that, with the statistics available at this moment, we see no indication of the optical alignment being not ok within the claimed accuracy of 45-50mu.

\begin{figure}[pht]
\begin{center}
\parbox{0.47\picwi}{
\includegraphics[width=0.45\picwi]{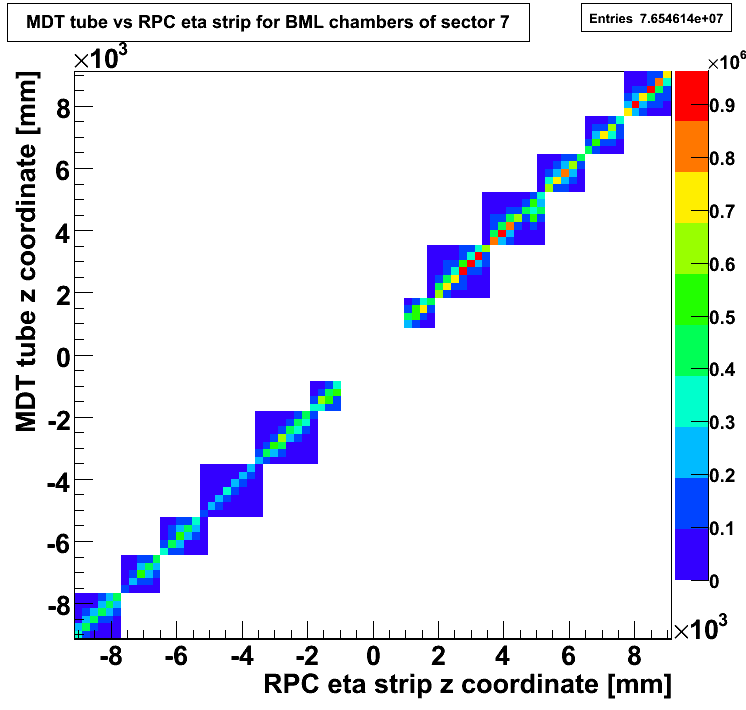}
\caption{\label{fig:mdt-rpc}Correlation of the $z$-coordinate of RPC hits (horizontal axis) and MDT hits (vertical axis) for cosmic muons.}
}
\parbox{0.47\picwi}{
\includegraphics[width=0.45\picwi]{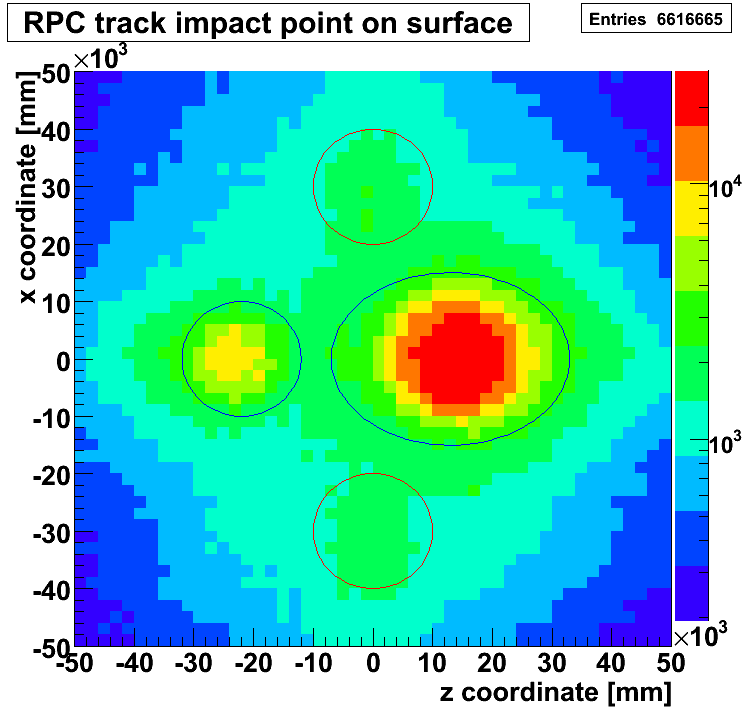}
\caption{\label{fig:rpc} Map of cosmic muons tracks projected onto the cavern surface.}
}
\end{center}
\end{figure}
In Figure\,\ref{fig:mdt-tgc}, a good correlation between MDT tube hits in the middle station and the TGC tracks interpolated to the MDT middle station is seen for cosmic events.
Vertical lines are due to noisy MDT tube while horizontal lines are due to noisy TGC wires.
%Correlation btw MDT tube hits in middle station and TGC track interpolated to MDT middle station for A side on run 91060. TGC tracks are reconstructed with wire hits and required at least 3 station hits.
A similar correlation is seen in Figure\,\ref{fig:mdt-rpc}, which shows the $z$-coordinates of the MDT tubes versus the RPC '$\eta$ strips' for cosmic muons.
% Spatial correlation between MDT tubes and RPC eta hits with cosmic data.The picture shows the scatter plot between MDT tubes z coordinate and RPC eta hits z coordinate of the same chamber along sector 7 for the Middle chamber. There is no clean-up cuts but an ADC cut > 50 on MDT Front-end response to reject MDT random hits. The blue squares are due to uncorrelated hits and show the station geometrical boundary along the z-axis.
Figure\,\ref{fig:rpc} projects cosmic muon tracks, reconstructed by the RPC, onto the cavern surface.
The two main access shafts -- the right one larger in diameter than the left one -- are clearly visible at $x=0$, as well as the two elevator shafts at $z=0$.
%
%Cosmics muon map reconstructed by off-line RPC standalone muon monitoring projected on surface (y=81m).The tracking is based on RPC space points, which are defined by orthogonal RPC cluster hits. The pattern recognition is seeded by a straight line defined by two space points belonging to the two Middle planes. Space points not part of any previous track and inside a predefined distance from the straight line are associated to the pattern. Patterns with points in at least 3 out of 4 layers in Middle planes are retained and a linear interpolation is performed in the two orthogonal views.

\section{Trigger and Data Acquisition}
The ATLAS Level-1 Trigger system consists of custom-built hardware that allows to form a trigger decision every 25\,ns based on information from the muon trigger detectors and the calorimeters.
The system is completely installed and rate tests proved successful readout of the whole ATLAS detector up to 40\,kHz, which is to be improved to the nominal rate of 75\,kHz in the summer of 2009.
The High-Level Trigger -- the Level-2 Trigger and Event Filter -- runs on CPU farms of currently 850\,PCs in 27 racks, which can either be used as Level-2 or Event Filter and is capable of sustaining a Level-1 rate of up to 60\,kHz.
The final configuration foresees 500\,PCs for Level-2 and 1800 for the Event Filter, with 8 cores each, running at 2.5\,GHz with 2\,GB RAM per core.
The finalisation of the system will be driven by the evolution of the LHC luminosity.
During cosmics data-taking, High-Level Trigger algorithms (e.g. tracking algorithms) have been successfully used to enrich cosmics samples for inner detector studies.

\section{Combined system}
\begin{figure}[pht]
\begin{center}
\includegraphics[width=0.5\picwi]{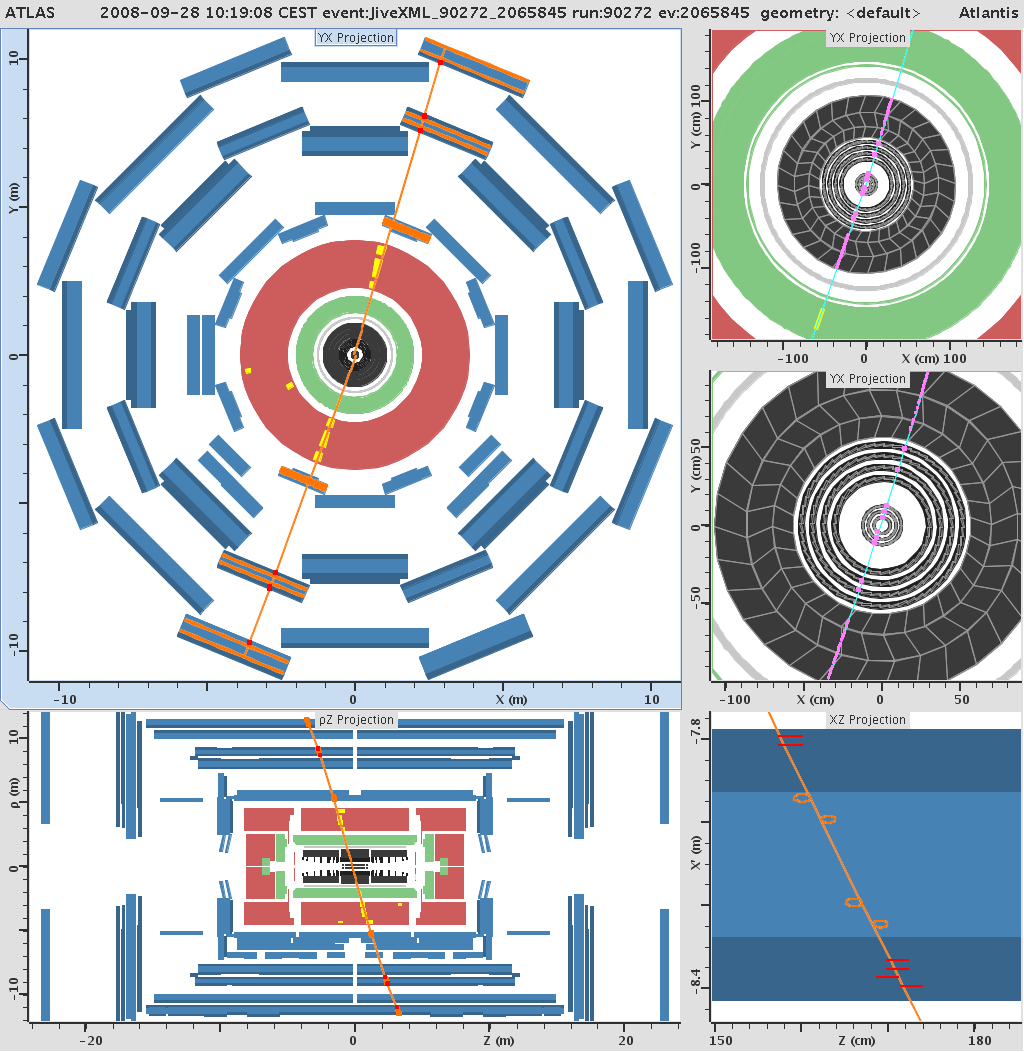}
\caption{\label{fig:combined-event}Event display of a cosmic muon leaving hits in all barrel detectors.}
\end{center}
\end{figure}

In this section, a few examples illustrate how the sub-detectors work together as a combined system.
Figure\,\ref{fig:combined-event} is an event display of a cosmic muon with hits in all barrel detectors, during a run when the solenoid and toroid magnets were in operation.

\newpage

\begin{figure}[pht]
\begin{center}
\parbox{0.47\picwi}{
\includegraphics[width=0.45\picwi]{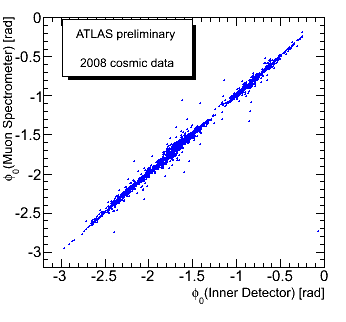}
\caption{\label{fig:combined-tracking-phi}Correlation of the azimuthal angle of cosmic muon tracks reconstructed by the inner detector and by the muon spectrometer.}
}
\parbox{0.47\picwi}{
\includegraphics[width=0.45\picwi]{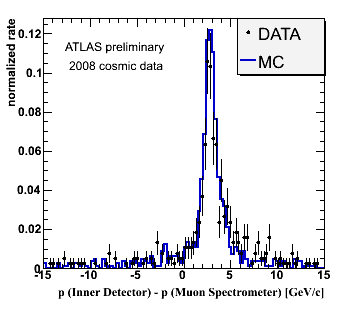}
\caption{\label{fig:combined-tracking-p}Difference in momentum of cosmic muon tracks reconstructed by the inner detector and by the muon spectrometer.}
}
\end{center}
\end{figure}
In Figure\,\ref{fig:combined-tracking-phi} and \ref{fig:combined-tracking-p} we see, for cosmics data, a comparison between quantities reconstructed for inner detector tracks and for matching muon spectrometer tracks.
A good correlation of all track parameters is seen:  the observed differences are well in agreement with the expected multiple scattering and energy loss in the calorimeters and are well reproduced by the Monte Carlo simulation.
This is illustrated in Figure\,\ref{fig:combined-tracking-phi}, in which the azimuthal coordinate $\phi_0$ is shown.
Figure\,\ref{fig:combined-tracking-p}, on the other hand, shows the difference in momentum, which is well within the expectation of a mean energy loss of 3\,GeV in the calorimeters.
No correction has been yet applied for the relative alignment of the inner detector and the muon spectrometer.

\section{Summary}
Commissioning of the ATLAS detector has started in 2005 along with the installation of the detector.
Large amounts of cosmics data have been taken in 2008 with all sub-detectors included.
In addition, ATLAS has been successfully operated with single beams from the LHC in September 2008.
Both data -- cosmics and beam data -- have been very useful for commissioning the detector, in particular for alignment and calibrations.
We have shown with a few examples that the ATLAS experiment is in good shape and ready for receiving proton-proton collisions at high energy and luminosity.

%\section*{Acknowledgments}

\end{document}